\documentclass{ws-p8-50x6-00}
\usepackage{epsfig}
\usepackage{color}
\usepackage{amsmath}
\usepackage{amssymb}

\def\iras{{\it IRAS}}
\def\pscz{{\it PSC}z}

\def\los{{\it l.o.s.}}
\def\gf{{\it g.f.}}

\begin{document}

\title{Large Scale Structure and Cosmic Rays revisited}

\author{ R. Ugoccioni, L. Teodoro and U. Wichoski}
\address{Centro Multidisciplinar de Astrof\'{\i}sica, Instituto Superior T\'ecnico,\\
 Av. Rovisco Pais 1, 1049-001 Lisboa, Portugal\\
roberto@fisica.ist.utl.pt, lteodoro@glencoe.ist.utl.pt and wichoski@ams3.ist.utl.pt}

%%%%%%%%%%%%%%%%%%%%%%%%%%%%%%%%%%%%%%%%%%%%%%%%%%%%%%%%%%%%%

\maketitle

\abstracts{
We investigate the possibility that ultra high energy 
cosmic rays ($E>10^{19}$ eV) are related 
to the distribution of matter on large scales. 
The large scale structure (LSS) data 
stems from the recent \iras\ \pscz\ redshift survey. 
We present preliminary predictions drawn from   an 
anisotropic distribution of sources which follows the galaxy distribution.
}

\section{Introduction}
\vspace{-8cm}
\begin{flushright}FISIST/12-2000/CENTRA
\end{flushright}
\vspace{8cm}
Ultra high energy cosmic rays (UHECR) are particles with kinetic energies 
above $\sim 10^{18}$ eV.\cite{nagano} The nature of these energetic 
particles is presently unknown. The reason is twofold: 
{\it i)} These particles interact on the top of the Earth's 
atmosphere producing extensive air 
showers (EAS) that can be observed from the ground; in this case, 
the primary particle can not be observed directly; and {\it ii)} 
at these ultra high energies (UHE) 
the fluxes are extremely low (less than 1 particle per square kilometer 
per year). This makes it impracticable to observe the UHE 
particles directly using balloons, satellites or spacecrafts 
due to their small acceptance.

The observation methods are indirect and rely on the 
observation of the secondary particles produced in the EAS. 
The hadronic particles, as well as muons and 
electrons created by the interactions of the primary particle 
in the atmosphere, 
are detected on the ground. The EAS also produces detectable 
fluorescent light 
photons due to the excitation of nitrogen molecules in the air by the 
charged secondary particles. Yet, the secondary charged particles 
that travel with velocities higher than the velocity of the light 
in the air generate Cherenkov radiation that can also be detected. 
Despite the fact that the EAS can be observed by the detection 
of different kinds of secondaries using various techniques, 
the determination of the nature of the
primary particle is very difficult and model dependent. 
As the fluxes are low it is necessary to use large ground 
arrays, and/or many fluorescent light and Cherenkov 
radiation detectors. 
Until the present moment, the ground arrays and the 
fluorescent light detectors have  
gathered only a handful of events in the UHE range. 
The number of events has not been enough to tell us whether  
the sources are extragalactic or are located in the Galaxy.

If the primary particle is a $\gamma$-ray or a neutrino the 
arrival direction would point back directly to the source. This 
would also be the case for charged particles if the Galactic magnetic 
field is $\lesssim 10^{-6}$ G and the
extragalactic magnetic fields are $\lesssim 10^{-9}$ G in the case of 
extragalactic sources. 
The distance to the source is also constrained for most kinds of 
primaries: If the primary particle is a nucleus or a proton 
(antiproton) the distance to the source is limited to less 
than $\sim 100$ Mpc for particles with arrival energies 
above $\sim 6 \times 10^{19}$ eV (GZK cutoff, see Fig.(1)). 
Due to interactions with the cosmic microwave background 
(CMB) photons these 
particles rapidly lose energy. Sources of $\gamma$-rays 
must be even closer because of the short mean absorption length 
for the UHE photons traveling in the CMB. 

The mechanism that provides particles with UHE is also not known. 
The ignorance about the sources makes it harder to 
determine the mechanism at work. 
For the UHE events no 
source candidate in the vicinity of the region to 
where the arrival direction points back has been found yet. 
On the other hand, some analysis of the 
showers profile have been favoring protons as the primary 
particle.\cite{ave} 
As it was mentioned above, the number of UHE events  
is still too small to allow us, based on the statistics, 
to answer questions about their isotropy and composition. 

In this work, we  assume that the UHECR primary particles 
above $10^{19}$ eV are 
predominantly extragalactic protons and that the sources 
are related to the distribution of matter on large scales. 
It means that without specifying the sources themselves or the 
acceleration mechanism, we would expect an excess of events coming 
from regions with mass overdensities and less events coming from 
regions with mass underdensities.
In the \S\ 2 we briefly describe the formalism used; the propagation 
code is described in the \S\ 3; and the  smoothing procedure of the 
density field is described in \S\ \ref{density_field}. Our results 
are presented in the \S\ 5.

\section{Formalism}
\label{sec:formalism}
In this contribution we apply a generalization of the formalism described in Waxman, Fisher and
Piran.\cite{waxman}
We model the population of UHECR sources $S$ 
within a ``box'' $\Delta V$ centered at $(z,\hat{\Omega})$ as drawn from a Poisson distribution
\[
\mbox{prob}_S(z,\hat{\Omega})=\frac{\bar{S}(z,\hat{\Omega})^S}{S!} \exp \left [ -\bar{S}
(z,\hat{\Omega}) \right ],
\label{eq:s_dist}
\]
whose mean value is $\bar{S}(z,\hat{\Omega})=\bar{s}(z)\frak{B}\left[\delta \rho(z,\hat{\Omega})\right]\Delta V$.
Here $\bar{s}(z)$ denotes the average comoving number of UHECR sources at redshift $z$ and 
$\frak{B}$ is some bias functional of the local galaxy distribution $\delta \rho(z,\hat{\Omega})$.
The generating function (\gf) of such a distribution is 
\begin{equation}
f_S(u;z,\hat{\Omega})=\exp\left[\bar{S}(z,\hat{\Omega})(u-1)\right],
\label{eq:gen_fun_S}
\end{equation}
where $u$ is a dummy variable. 

The detected number $N$ of UHECR  produced by a source within 
$\Delta V$ with observed energy larger than $E$ is also modeled by a Poisson distribution 
\[
\mbox{prob}_N( \ge E)=
\frac{\bar{N}(E,z)^N}{N!} \exp \left \{-\bar{N}(E,z) \right \},
\label{eq:n_ener}
\]
with mean value
\[
\bar{N}(E,z) = {\cal{A\,T}}\frac{\dot{n}_0\left[E_{inj}(E,z)\right]}{\bar{s}_0} \frac{(1+z)}{4\pi d_L(z)^2},
\label{eq:n_mean}
\]
where 
$\bar{s}_0 = \bar{s}(z = 0)$, $d_L^2(z)=4c^2 H^{-2}_0(2+z-2\sqrt{1+z})$ %~\cite{Kolb:1990}
for an $\Omega = 1$ Universe; $\cal{A}$ and $\cal{T}$ denote the detector area and observation
time, respectively; $E_{inj}$ is the energy with which a UHECR observed with 
energy $E$ was produced at redshift $z$; and 
$\dot n_0$ is the number of UHE protons emitted by a source per unit 
time and is assumed to be proportional to $dN/dE_{inj}$. 
We have assumed that the source differential spectrum is a power law 
in energy $ dN/dE_{inj} \propto E_{inj}^{-(\gamma+1)} $.
The \gf\ of the last probability distribution is given by
\begin{equation}
g_N (u;z,E)= \exp\left[ \bar{N}(E,z)(u-1)\right].
\label{eq:gen_fun_N}
\end{equation}
Hence, it is straightforward to show from equations (\ref{eq:gen_fun_S}) and  
(\ref{eq:gen_fun_N}) that the \gf\ for the probability of observing a total of $N$\ events 
from $\Delta V$, with an energy larger than $E$, is expressed by
\[
F(u;z,\hat{\Omega},E)= \exp\left\{\bar{S}(z,\hat{\Omega})
\left[\exp\left(\bar{N}(E,z)(u-1)\right)-1\right] \right \}.
\label{eq:gen_fun_NS}
\]
The overall UHECR distribution coming from a collection of independent volume elements 
$\Delta V_i$ has the 
following \gf:
\begin{equation*}
F(u;\bigcup_i \Delta V_i,E)=\prod_i F\left(u;z_i,\hat{\Omega}_i,E\right), 
\end{equation*}
which for a given line of sight (\los) can be expressed as an integral over $z$
\begin{equation}
F(u;\hat{\Omega},E)=\exp\left\{\int_0^{z_{max}} \bar{S}(z,\hat{\Omega}) \left[\exp\left(\bar{N}(E,z)(u-1)\right)-1\right] dV\right\},
\label{eq:gen_fun_NS_los}
\end{equation}
where $dV = c\,\left | dt/dz \right | d_L^2(z)(1+z)^{-1}\,dz$.

In defining $\lambda(\hat{\Omega})$ as 
\[
\lambda(\hat{\Omega})\equiv\int _0^{z_{max}}\bar{S}(z,\hat{\Omega})dV,
\]
one can characterize the distribution of UHECR produced by sources along the \los\ with 
energy larger than $E$ by the \gf.
\[
G(u;\hat{\Omega},E) \equiv \frac{1}{ \lambda(\hat{\Omega})} \int_{0}^{z_{max}} \bar{S}(z,\hat{\Omega})\exp\left(\bar{N}(E,z)(u-1)\right) dV.
\]
From equation (\ref{eq:gen_fun_NS_los}) is then straightforward to show that 
\[
F(u;\hat{\Omega},E)=\exp\left\{\lambda(\hat{\Omega})\left[G(u;\hat{\Omega},E)-1\right]\right\},
\]
which still is a compound Poisson distribution, although $G$ is not Poissonian.

\section{Propagation code}

The propagation equation takes 
into account energy losses of the UHE protons due 
to: {\it i)} Adiabatic expansion of the Universe; 
{\it ii)} $e^+ e^-$ pair 
production; and {\it iii)} pion production due 
to interactions with CMB photons.
A proton observed at present 
($z=0$) with energy $E$ must have been produced at an epoch $z$ 
with energy $E_{inj} = E_{inj}(E,z)$. We assume that 
the influence of the magnetic fields 
on particles with energies $E > 10^{19}$ eV is negligible. Figure 1 
shows the decrease of energy as a function of the distance from the source for 
UHE protons. We note that for a proton to be observed with energies above 
$\sim 6 \times 10^{19}$ eV the source must be within 
$\sim 100$ Mpc from the observer irrespective to $E_{inj}$. 

\begin{figure}
\begin{center}
\scalebox{0.35}{\epsfig{angle=-90,file=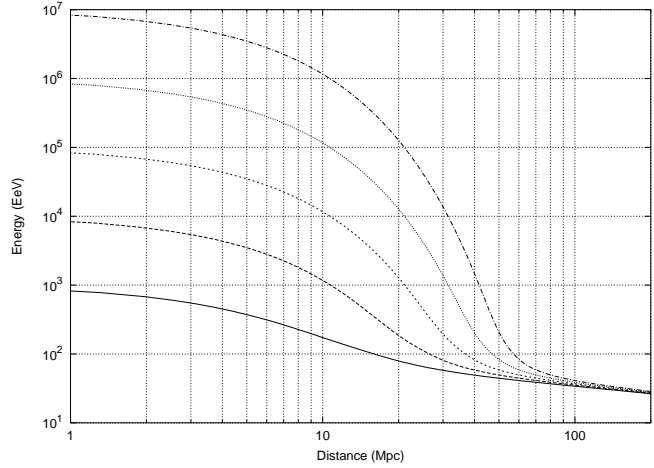}}
\end{center}
\caption{Propagation of UHE protons in the CMB. The lines represent 
various proton injection energies, $E_{inj} = 10^{7}, 10^{6}, 10^{5},
10^{4}, 10^{3}$ EeV (from top to bottom respectively).}
\end{figure}

\section{Smoothed density field}
\label{density_field}

The galaxy distribution is estimated from the \iras\ \pscz\ redshift 
survey.\cite{Saunders:2000}
We have computed the smoothed density field on a spherical grid up to
200\,$h^{-1}$\,Mpc. The Gaussian-smoothed density field at a grid point $n$ is given by
\begin{equation}
\label{smooth_dens_field}
1+\delta_g(c\vec z_n)= \frac{1}{(2\pi) ^{3/2}\sigma_{sm,n}^3}\sum _i \frac{1}{\phi(c{\vec z} _i)}\exp \left [ -\frac{(c\vec z _n -c\vec z_i)^2}{2\sigma ^2_{sm,n}} \right ].
\end{equation}
We have divided the sphere in 72 bins of approximately equal area.
 Radially, the bin size increases in  proportion to the \iras\ \pscz\ 
inter-particle spacing $[\bar{n}\phi(cz)]^{-1/3}$.  This smoothing scheme is tailored to keep the
shot-noise uncertainties in the density field roughly constant through out the sampled  volume. A 
more detailed analysis of the \iras\ \pscz\ density field can be found in Branchini {\it et al.}\cite{Branchini:1999b}

\section{Results}
\label{subs:results}

We have found that the final results are independent of the cosmological parameters 
$(\Omega,\Lambda)$. Thus, we have used $ \Omega =1 $ throughout our calculations for the 
sake of simplicity. For the bias functional $\frak{B} [\delta(\vec x)]$\ we have considered $\frak{B}  [\delta(\vec x)]= 1+\delta (\vec x)$.
\begin{figure*}
\centering
\scalebox{0.60}{\includegraphics[30,350][588,520]{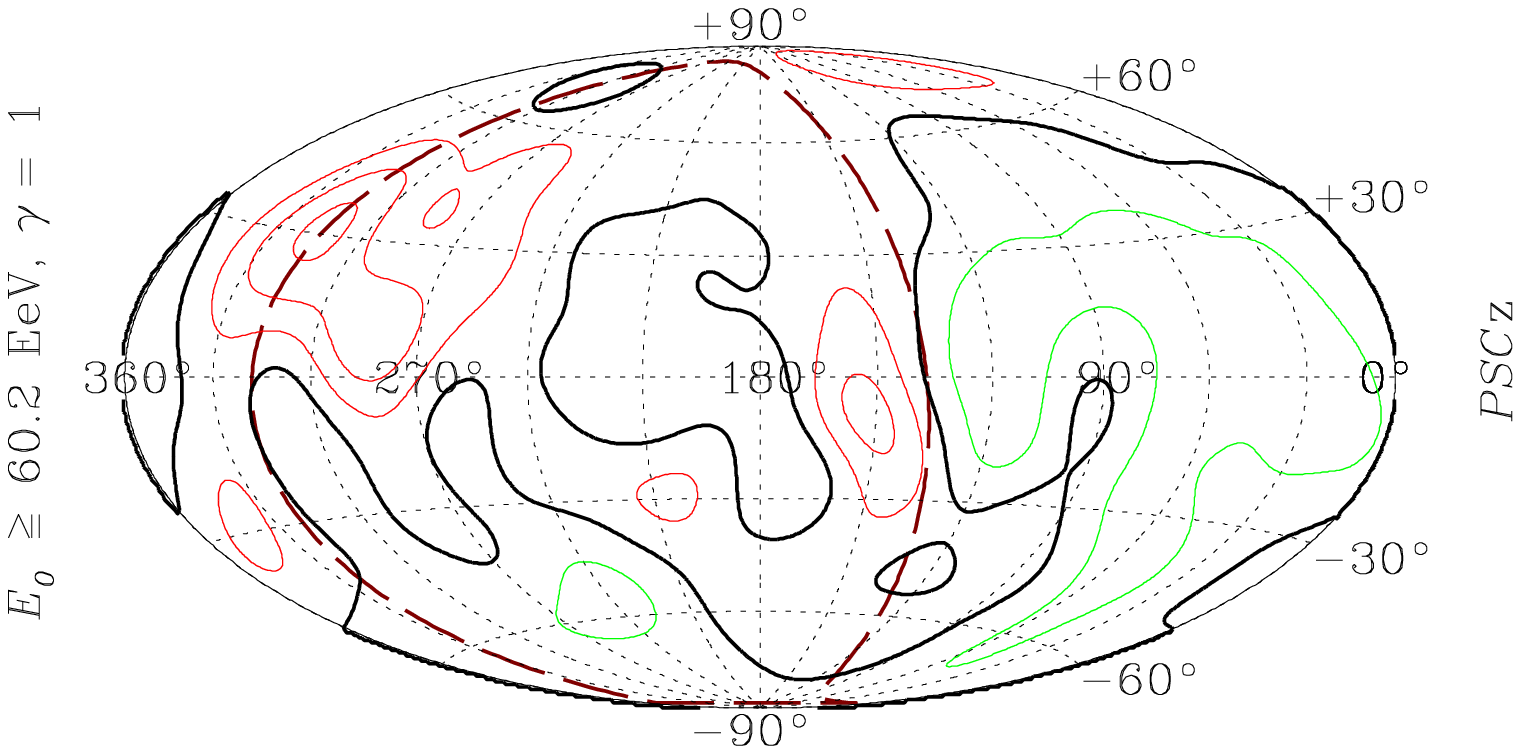}}
\scalebox{0.60}{\includegraphics[30,300][588,570]{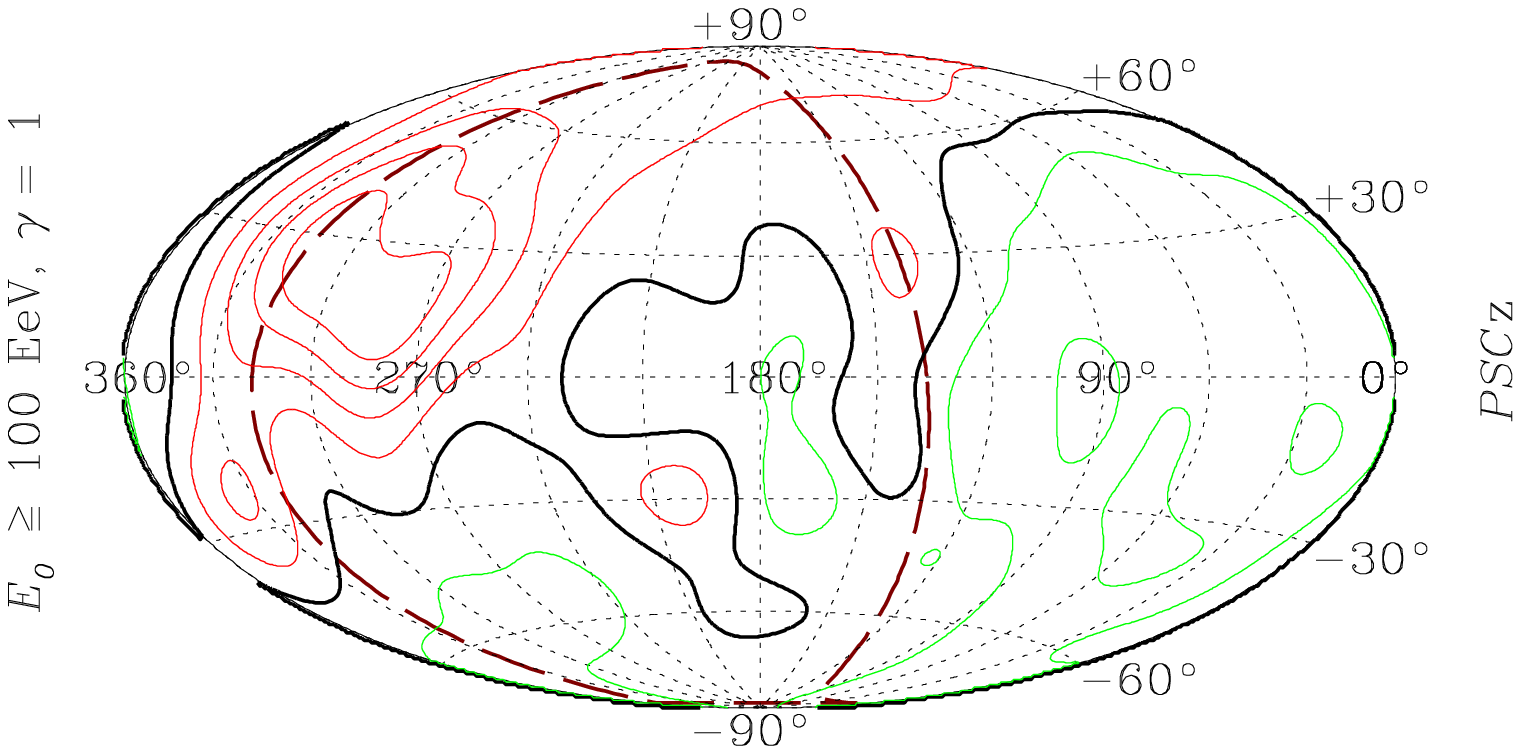}}
\caption{Aitoff projection of $\delta_{CR}$, for {\it E}$_{inj}  \approx$ 
60 and 100 EeV ($\gamma = 1.0$). 
The heavy contour denotes the zero contour. Dark (light) grey 
contours denote positive (negative) fluctuations equally-spaced at 0.20. The 
long-dashed line represents the Super-Galactic plane.} 
\label{fig:sky}
\end{figure*}
Figure \ref{fig:sky} presents maps of fluctuations in the mean Cosmic Ray intensity, 
\begin{equation}
\delta_{CR}(E,\hat{\Omega})= \dfrac{ 4\pi \bar N(E,\hat{\Omega})}{\int d \hat \Omega 
\bar N (E,\hat{\Omega})} -1, 
\end{equation}
for $E = (6,10) \times 10^{19}$~eV. 
In the maps we clearly  see the regions from where an 
excess and a deficit of UHECR events is expected following the LSS. 
The specific predictions for future experiments as the 
Auger project and HiRes will be presented elsewhere.\cite{lru}

\vspace{0.20cm}
\noindent
{\bf Acknowledgements:}
This work has been supported by ``Funda\c{c}\~ao para a Ci\^encia e a 
Tecnologia'' (FCT) under the program ``PRAXIS  XXI''.  L.T. has also been 
supported by FCT under the project PRAXIS/C/FIS/13196/98.

\end{document}